\documentclass[onecolumn,prd,amsmath,amssymb,showkeys]{revtex4}

\usepackage{amsmath}
\usepackage{graphicx}
\usepackage{graphics}
\usepackage{dcolumn}
\usepackage{bm}
\usepackage{hyperref}

\def\BEq{\begin{equation}}
\def\EEq{\end{equation}}
\def\BEqA{\begin{eqnarray}}
\def\EEqA{\end{eqnarray}}
\def\BEn{\begin{enumerate}}
\def\EEn{\end{enumerate}}
\def\BWT{\begin{widetext}}
\def\EWT{\end{widetext}}

\def\a{\alpha}

\def\d{\delta}

\def\g{\gamma}

\def\dag{\dagger}
\def\adj{{^{\dag}}}

\def\bra{\langle}

\def\ket{\rangle}

\newcommand{\cev}[1]{\reflectbox{\ensuremath{\vec{\reflectbox{\ensuremath{#1}}}}}}

\begin{document}

\title{Positive-energy Dirac particles in cosmology}

\author{Andrei Galiautdinov}
\email{ag1@uga.edu}
 \affiliation{
Department of Physics and Astronomy, 
University of Georgia, Athens, GA 30602, USA}

\date{\today}
\begin{abstract}

We provide general relativistic treatment of 
the single-component field described by Dirac's positive-energy
wave equation of 1971. It is motivated by Bogomolny's proposal 
to regard that field as a possible candidate for dark matter. Our emphasis 
is on standard (flat) Friedmann-Robertson-Walker cosmology. 
The early universe is not considered, though there is a strong indication 
that the field may play a role in internally driven generation of cosmic
anisotropy.
\end{abstract}


\maketitle


\section{Introduction}

The discussion centers around the positive-energy relativistic 
wave equation (colloquially known as the new Dirac equation),
\BEq
\label{eq:NDE}
({\gamma}^a \partial_a - m) q\psi = 0,
\EEq
where
\BEq
q
=
\begin{pmatrix}
q_1 \cr
q_2 \cr
q_3 \cr
q_4
\end{pmatrix},
\quad
q\psi
=
\begin{pmatrix}
q_1 \psi \cr
q_2 \psi \cr
q_3 \psi \cr
q_4 \psi
\end{pmatrix},
\quad
{\gamma}^a {\gamma}^b 
+ {\gamma}^b {\gamma}^a = -2\eta^{ab},
\quad
\eta^{ab} = {\rm diag}(+,-,-,-),
\quad
a,b=0,1,2,3,
\EEq 
put forward by Paul Dirac in 1971 
\cite{dirac1971,dirac1972,dirac1978}.
Being superficially similar in appearance to his famous 
spinorial equation of 1928 \cite{dirac1928}, the new 
equation describes a {\it single-component} field, 
$\psi(x; q_1,q_2)$, which, in addition to the usual 
spacetime coordinates, $x=(x^a)$, also depends on 
a pair of continuous real-valued variables, $q_{1}$ and 
$q_{2}$, representing two internal quantum-mechanical 
degrees of freedom. A simple way to visualize this field, 
which slightly deviates from Dirac's original single-particle 
interpretation, is to imagine that at every spacetime point, 
$P=(x^0,x^1,x^2,x^3)$, there exists an intrinsic 
infinite-dimensional Hilbert space ${\cal H}$ of 
internal quantum states, $|\psi\ket$, on which there operate 
two internal ``position'' operators, $\hat{q}_1$ and 
$\hat{q}_2$ (denoted with an  overhat). These operators 
form a complete set of commuting internal observables and 
satisfy the eigenvalue problem,
\BEq
\hat{q}_1 |q_1,q_2\ket = q_1 |q_1,q_2\ket,
\quad
\hat{q}_2 |q_1,q_2\ket = q_2 |q_1,q_2\ket,
\quad
q_1, q_2 \in \mathbb{R}.
\EEq 
The corresponding quantum amplitude ${\cal A}$ describing 
internal transition from some initial state $|\psi\ket$ to the 
final state $\bra q_1,q_2|$ (which is presumably going on in 
Nature without experimenter's direct involvement) is then 
represented by the internal wave function,
\BEq
{\cal A} = \bra q_1,q_2|\psi\ket \equiv \psi(q_1,q_2).
\EEq 
Since $|\psi\ket$ may be chosen arbitrarily at every $P$, 
the amplitude acquires a spacetime label which turns it into 
a field,
\BEq
{\cal A}(x) = \bra q_1,q_2|\psi(x)\ket \equiv \psi(x; q_1,q_2).
\EEq 
In the internal ``position'' representation, the action of 
$\hat{q}_{1,2}$ on the internal wave function $\psi(q_1,q_2)$ 
is naturally implemented by simple multiplication, 
\BEq
\bra q_1,q_2|\hat{q}_j|\psi\ket = q_j \psi(q_1,q_2), 
\quad
j = 1,2,
\EEq
so that $\hat{q}_{1,2} \rightarrow q_{1,2}$, hence the 
ubiquitous appearance of products $q\psi$ in Dirac's  
theory. Notice that the  corresponding internal ``momentum'' 
operators, $\hat{\pi}_1 \rightarrow -i\partial/\partial q_1 \equiv q_3$ 
and $\hat{\pi}_2 \rightarrow -i\partial/\partial q_2 \equiv q_4$, 
are implemented by differentiation, as usual, and, together with 
$\hat{q}_1$ and $\hat{q}_2$, obey the standard canonical 
commutation relations,
\BEq
[\hat{q}_1,\hat{q}_2]=[\hat{q}_1,\hat{\pi}_2]
=[\hat{q}_2,\hat{\pi}_1]=[\hat{\pi}_1,\hat{\pi}_2]=0,
\quad
[\hat{q}_1,\hat{\pi}_1]=[\hat{q}_2,\hat{\pi}_2]=i,
\EEq
or,
\BEq
[q_1,q_2]=[q_1,q_4]=[q_2,q_3]=[q_3,q_4]=0,
\quad
[q_1,q_3]=[q_2,q_4]=i,
\EEq
which can be summarized in matrix form (the tilde indicates 
matrix transposition),
\BEq
[q_k,q_\ell]=i (\gamma^0)_{k\ell},
\quad
k, \ell = 1,2,3,4,
\quad
\gamma^0
\equiv
\begin{pmatrix}
0 & 0 & 1 & 0 \cr
0 & 0 & 0 & 1 \cr
-1 & 0 & 0 & 0 \cr
0 & -1 & 0 & 0
\end{pmatrix},
\quad
(\gamma^0)^2 = -1,
\quad
\tilde{\gamma}^0=-\gamma^0.
\EEq
The antisymmetric $\gamma^0$ appearing here is the same 
as the one used in the actual equation (\ref{eq:NDE}); the 
remaining gammas, being real and symmetric, are chosen 
in the form, 
\BEq
\gamma^1
\equiv
\begin{pmatrix}
-1 & 0 & 0 & 0 \cr
0 & 1 & 0 & 0 \cr
0 & 0 & 1 & 0 \cr
0 & 0 & 0 & -1
\end{pmatrix},
\quad
\gamma^2
\equiv
\begin{pmatrix}
0 & 1 & 0 & 0 \cr
1 & 0 & 0 & 0 \cr
0 & 0 & 0 & -1 \cr
0 & 0 & -1 & 0
\end{pmatrix},
\quad
\gamma^3
\equiv
\begin{pmatrix}
0 & 0 & -1 & 0 \cr
0 & 0 & 0 & -1 \cr
-1 & 0 & 0 & 0 \cr
0 & -1 & 0 & 0
\end{pmatrix}.
\EEq

To arrive at Eq.\ (\ref{eq:NDE}), one introduces the four 
kets, $\hat{q}_1|\psi(x)\ket$, $\hat{q}_2|\psi(x)\ket$, 
$\hat{\pi}_1|\psi(x)\ket$, $\hat{\pi}_2|\psi(x)\ket$, forms 
the corresponding amplitudes representing internal transition 
processes, $\bra q_1,q_2|\hat{q}_1|\psi(x)\ket$, 
$\bra q_1,q_2|\hat{q}_2|\psi(x)\ket$, 
$\bra q_1,q_2|\hat{\pi}_1|\psi(x)\ket$, 
$\bra q_1,q_2|\hat{\pi}_2|\psi(x)\ket$, and then 
assembles the amplitudes into a four-dimensional column,
\BEq
\bra q_1,q_2|\hat{q}|\psi(x) \ket
\equiv
\begin{pmatrix}
\bra q_1,q_2|\hat{q}_1|\psi(x)\ket \cr
\bra q_1,q_2|\hat{q}_2|\psi(x)\ket \cr
\bra q_1,q_2|\hat{\pi}_1|\psi(x)\ket \cr
\bra q_1,q_2|\hat{\pi}_2|\psi(x)\ket
\end{pmatrix}
=
\begin{pmatrix}
q_1 \psi(x;q_1,q_2) \cr
q_2 \psi(x;q_1,q_2) \cr
q_3 \psi(x;q_1,q_2) \cr
q_4 \psi(x;q_1,q_2)
\end{pmatrix}
=
\begin{pmatrix}
q_1  \cr
q_2  \cr
q_3  \cr
q_4 
\end{pmatrix}
\psi(x;q_1,q_2)
\equiv
q\psi(x;q_1,q_2).
\EEq
Equation (\ref{eq:NDE}) may then be viewed as the 
postulate of the theory,
a law of Nature, which imposes certain nontrivial restrictions on 
how the four dynamically evolving transition amplitudes 
mentioned above must be related to each other.

Dirac's positive-energy field displays some quite unusual 
properties (summarized in the wonderful paper by 
Bogomolny \cite{bogomolny2024}), most interesting being 
the positivity of energy of its particles and the impossibility of 
coupling to electromagnetic field \cite{dirac1971}. The latter 
property signals impossibility of detection by regular optical means 
which, in turn, indicates that the new Dirac field may potentially 
play a role in cosmology as a possible candidate for dark matter \cite{bogomolny2024}.

\section{Curved spacetime: action, equations of motion, stress-energy tensor}

Working on a curved background, we postulate the material 
action for the dark Dirac field to be 
({\it cf}.\ \cite{ahner1975,ahner1976,castillo2011,ousmanemanga2013,birrell1984}; 
integration $\int dq_1 dq_2$ is implied),
\begin{align}
S_{\rm D}=\int d^4 x \; {\cal L}_{\rm D},
\quad
{\cal L}_{\rm D}
=
-i \, \sqrt{-g}
\left\{
\frac{1}{2}
\left[
\bar{\Psi} \breve{\gamma}^\mu (\nabla_\mu \Psi)
-
(\nabla_\mu \bar{\Psi}) \breve{\gamma}^\mu \Psi
\right]
-m \bar{\Psi} \Psi
\right\},
\end{align}
with
\BEq
\Psi = q\psi,
\quad
\Psi\adj = \psi\adj\tilde{q},
\quad
\bar{\Psi} = \Psi\adj \gamma^0 = \psi\adj\tilde{q} \gamma^0
=\psi\adj \bar{q},
\quad
\bar{q}=\tilde{q}\gamma^0,
\quad
\tilde{q}=(q_1, q_2, q_3, q_4),
\EEq
and
\BEq
\breve{\gamma}^\mu(x) = e^{\mu}_{a}(x){\gamma}^a,
\quad
\breve{\gamma}^\mu(x) \breve{\gamma}^\nu(x) 
+ \breve{\gamma}^\nu(x) \breve{\gamma}^\mu(x) = -2g^{\mu\nu}(x),
\quad
\mu,\nu = 0, 1,2,3,
\EEq
where $x=(x^\mu)$ are some arbitrary large-scale coordinates 
(in contrast to the local Minkowskian coordinates, $x^a$), 
$e^{\mu}_{a}=\partial x^\mu/\partial x^a$ is the tetrad (vierbein) 
field, $\sqrt{-g} = {\rm det}(e^{a}_{\mu})$, with breve indicating 
$x^\mu$-dependent gamma matrices. We regard $\bar{\Psi}$ 
and $\Psi$ as the independent dynamical variables of the theory 
whose covariant derivatives are given by,
\BEq
\nabla_{\nu}\Psi 
= (\partial_{\nu} + \Omega_{\nu})\Psi,
\quad
\nabla_{\nu}\bar{\Psi} 
= \bar{\Psi}(\cev{\partial}_{\nu} - \Omega_{\nu}),
\EEq
where
\BEq
\Omega_{\nu}
=
-\frac{1}{4}\omega_{ab\nu}(\sigma^{ab}+is^{ab}),
\EEq
with the standard spin connection coefficients,
\BEq
\omega_{ab\nu}
= 
e_{a\mu}(e^{\mu}_{b,\nu}
+e^{\sigma}_{b}\Gamma^{\mu}_{\sigma \nu}),
\EEq
and
\BEq
\sigma^{ab}= \frac{1}{2}[\gamma^a,\gamma^b],
\quad
s^{ab} = \frac{1}{2}(\bar{q}\sigma^{ab}q),
\quad
\bar{q} = \tilde{q}\gamma^0.
\EEq

Variation with respect to $\bar{\Psi}$ and $\Psi$ immediately 
recovers the new Dirac equation together with its adjoint, 
\BEq
(\breve{\gamma}^\mu \nabla_\mu - m) \Psi = 0,
\quad \bar{\Psi}(\cev{\nabla}_\mu \breve{\gamma}^\mu + m) = 0,
\EEq
while variation with respect to the metric, $g_{\mu\nu}$, results 
in the stress-energy tensor,
\BEq
T_{\mu\nu}
= 
-i \, 
\left(
\frac{1}{2}
\left[
\bar{\Psi} \breve{\gamma}_{(\mu} \nabla_{\nu)} \Psi
-
\nabla_{(\mu} \bar{\Psi} \breve{\gamma}_{\nu )} \Psi
\right]
-
\frac{1}{2}g_{\mu\nu}
\left[
\bar{\Psi} \breve{\gamma}^{\alpha} \nabla_{\alpha} \Psi
-
\nabla_{\alpha} \bar{\Psi} \breve{\gamma}^{\alpha} \Psi
\right]
+m g_{\mu\nu}\bar{\Psi} \Psi
\right) + \text{``divergence term''},
\EEq
whose calculation is given in Appendix \ref{appendix1}. 
In the classical limit the divergence term vanishes. 
On the mass shell the expression then reduces to
\BEq
T_{\mu\nu}
= 
\frac{-i}{2}
\left[
\bar{\Psi} \breve{\gamma}_{(\mu} \nabla_{\nu)} \Psi
-
\nabla_{(\mu} \bar{\Psi} \breve{\gamma}_{\nu )} \Psi
\right] \quad \text{(on-shell)},
\EEq
whose trace is
\BEq
\label{eq:T}
T 
\equiv 
T^{\mu}_{\mu} 
= 
-i \, m \bar{\Psi} {\Psi} 
=
-i \, m {\psi}\adj \bar{q}q \psi
=
2m \int dq_1dq_2 \; \psi\adj\psi
\quad \text{(on-shell)},
\EEq
where we used $\bar{q}q=\tilde{q}\gamma^0 q = 2i$.

We note in passing \cite{valenzuela2024}
that if the variation were performed with respect 
to the (little) $\psi$ then the resulting equation of motion would 
coincide with the positive-energy Majorana equation 
\cite{majorana1932} (also, \cite{horvathy2008,bekaert2009}),
\BEq
\label{eq:ME}
(\bar{q}{\gamma}^a q \nabla_a - m \bar{q}q) \psi = 0.
\EEq
In what follows we will adopt the action principle based on the 
``big'' $\Psi$, leaving the question as to the relationship between 
these two variational approaches to future study.
 
\section{Friedmann-Robertson-Walker cosmology as an example}

The explicit calculation is particularly simple when working with 
conformally-flat Friedmann-Robertson-Walker metric 
\cite{shapiro2022,galiautdinov2016}, 
$g_{\mu\nu}= a^2(\eta) \eta_{\mu\nu}$, where $a(\eta)$ 
is the cosmological scale factor, and $\eta$ is the conformal 
time, which is  related to the standard cosmological time, 
$t_{\rm H}$, measured by the freely falling Hubble observers 
via the differential relation,
\BEq
dt_{\rm H}= a(\eta)d\eta.
\EEq
In this case, the non-vanishing Christoffel symbols are given 
by (the prime indicates differentiation with respect to $\eta$),
\BEq
\Gamma^{0}_{00}=\frac{a'}{a},
\quad
\Gamma^{0}_{ij}=\frac{a'}{a}\delta_{ij},
\quad
\Gamma^{i}_{0j}=\frac{a'}{a}\delta^{i}_{j},
\quad
i,j=1,2,3,
\EEq
where $a'/a$ should be interpreted as conformal Hubble parameter. 
The tetrad fields are,
\BEq
e^{b}_{\mu}=a\delta^{b}_{\mu},
\quad
e^{\mu}_{b}=\frac{1}{a}\delta^{\mu}_{b},
\EEq
resulting in the spin connection,
\BEq
\omega_{011}=\omega_{022}=\omega_{033}=-\omega_{101}
=-\omega_{202}=-\omega_{303}=\frac{a'}{a},
\EEq
the gamma matrices, 
\BEq
\breve{\gamma}^{0}=\frac{1}{a}\gamma^{0},
\quad
\breve{\gamma}_{0}=a\gamma^{0},
\quad
\breve{\gamma}^{k}=\frac{1}{a}\gamma^{k},
\quad
\breve{\gamma}_{k}=-a\gamma^{k}=a\gamma_{k},
\quad k = 1,2,3,
\EEq
and connection for the field,
\BEq
\Omega_0=0,
\quad
\Omega_k 
= 
-\frac{a'}{2a}(\sigma^{0k} + i s^{0k})
=
-\frac{a'}{2a}(\gamma^0\gamma^k - i \tilde{q}\gamma^k q).
\EEq
From the equations of motion we also find the time-derivatives 
of the (here position-independent) fields,
\BEq
\Psi' = \g^0\g^k\Omega_k \Psi - a m \g^0 \Psi,
\quad
\bar{\Psi}'=-\bar{\Psi}\Omega_k\gamma^k\gamma^0+am\bar{\Psi}\gamma^0.
\EEq

Calculation of the on-shell stress-energy tensor proceeds as follows:

For the 00-component we have,
\begin{align}
\label{eq:T00calculation}
T_{00} & =
-\frac{i}{2}
\left[\bar{\Psi}\breve{\g}_0 \Psi' -\bar{\Psi}'\breve{\g}_0 \Psi\right]
 =
-\frac{i}{2}g_{00}
\left[\bar{\Psi}\breve{\g}^0 \Psi' -\bar{\Psi}'\breve{\g}^0 \Psi\right]
 =
-\frac{ia}{2} 
\left[\bar{\Psi}{\g}^0 \Psi' -\bar{\Psi}'{\g}^0 \Psi\right]
\nonumber \\
&=
-\frac{ia}{2}
\left[2a m\bar{\Psi}\Psi 
+ \bar{\Psi}({\g}^k\Omega_k + \Omega_k \gamma^k)\Psi\right]
\nonumber \\
& =
-ia^2m \bar{\Psi}\Psi
+\frac{ia'}{4}
\left[\bar{\Psi}({\g}^k (\sigma^{0k} + i s^{0k})
 + (\sigma^{0k} + i s^{0k}) \gamma^k)\Psi\right]
\nonumber \\
&=
2a^2m {\psi}\adj\psi
-\frac{a'}{4}
{\psi}\adj 
\left\{
\bar{q}
\left({\g}^k s^{0k} +s^{0k}\gamma^k\right) q
\right\}
\psi.
\end{align}
Now notice that,
\begin{align}
\bar{q} \left({\g}^1 A + A \gamma^1\right) q
&= 2(q_1 A q_3 + q_3 A q_1) - 2(q_2 A q_4 + q_4 A q_2),
\\
\bar{q} \left({\g}^2 B + B \gamma^2\right) q
&= -2(q_1 B q_4 + q_4 B q_1) - 2(q_2 B q_3 + q_3 B q_2),
\\
\bar{q} \left({\g}^3 C + C \gamma^3\right) q
&= -2(q_1 C q_1 + q_2 C q_2) +2 (q_3 C q_3 + q_4 C q_4).
\end{align}
and since in this case
\begin{align}
A &= s^{01} = q_1^2 - q_2^2 - q_3^2 + q_4^2,
\\
B &= s^{02} = - q_1 q_2 - q_2 q_1 + q_3 q_4 + q_4 q_3,
\\
C&= s^{03} = q_1 q_3 + q_3 q_1 + q_2 q_4 + q_4 q_2,
\end{align}
the combination in curly brackets in (\ref{eq:T00calculation}) vanishes, and we get,
\begin{align}
\label{eq:T00}
T_{00} 
&=
-ia^2m \bar{\Psi}\Psi = 2a^2m \int dq_1 dq_2 {\psi}\adj\psi.
\end{align}
Comparison with (\ref{eq:T}) immediately leads to the conclusion that
\BEq
T_{11}+T_{22}+T_{33}=0,
\EEq
which can be verified by direct calculation, with the result,
\begin{align}
\label{eq:T11calculation}
T_{11} 
&=
\frac{a'}{4}
{\psi}\adj \bar{q}
\left({\g}^1 s^{01} + s^{01} \gamma^1\right) q\psi
=
\frac{a'}{4}
\int dq_1dq_2
{\psi}\adj 
\left(4(q_1q_3 - q_2q_4)(q_1^2 - q_2^2 - q_3^2 + q_4^2)\right) 
\psi,
\\
\label{eq:T22calculation}
T_{22} 
&=
\frac{a'}{4}
{\psi}\adj \bar{q}
\left({\g}^2 s^{02} + s^{02} \gamma^2\right) q\psi
=
\frac{a'}{4}
\int dq_1dq_2
{\psi}\adj 
\left(8(q_1q_4 + q_2q_3)(q_1q_2 - q_3q_4)\right) 
\psi,
\\
\label{eq:T33calculation}
T_{33} 
&=
\frac{a'}{4}
{\psi}\adj \bar{q}
\left({\g}^3 s^{03} + s^{03} \gamma^3\right) q\psi
=
\frac{a'}{4}
\int dq_1dq_2
{\psi}\adj 
\left(-4(q_1q_3 + q_2q_4)(q_1^2 + q_2^2 - q_3^2 - q_4^2)\right) 
\psi,
\end{align}
indicating that we are dealing with quite unusual anisotropic essence. 
As to the other components of the stress-energy tensor, we find,
\begin{align}
\label{eq:T01calculation}
T_{01} 
& =
-\frac{i}{4}
\left[
\bar{\Psi}\breve{\g}_0 \nabla_1 \Psi 
+ \bar{\Psi}\breve{\g}_1 \nabla_0 \Psi 
- (\nabla_0\bar{\Psi})\breve{\g}_1 \Psi
- (\nabla_1\bar{\Psi})\breve{\g}_0 \Psi\right]
\nonumber \\
& =
-\frac{i}{4}
\left[
g_{00}\bar{\Psi}\breve{\g}^0 \nabla_1 \Psi 
+ g_{11}\bar{\Psi}\breve{\g}^1 \nabla_0 \Psi 
- g_{11}(\nabla_0\bar{\Psi})\breve{\g}^1 \Psi
- g_{00}(\nabla_1\bar{\Psi})\breve{\g}^0 \Psi\right]
\nonumber \\
& =
-\frac{ia}{4}
\left[
\bar{\Psi}({\g}^0 \Omega_1 + \Omega_1 {\g}^0) \Psi 
-\bar{\Psi}{\g}^1  \Psi'
+\bar{\Psi}'{\g}^1 \Psi
\right]
\nonumber \\
& =
-\frac{ia}{4}
\left[
\bar{\Psi}({\g}^0 \Omega_1 + \Omega_1 {\g}^0) \Psi 
-\bar{\Psi}{\g}^1 (\g^0\g^k\Omega_k \Psi - a m \g^0 \Psi)
+(-\bar{\Psi}\Omega_k\gamma^k\gamma^0+am\bar{\Psi}\gamma^0){\g}^1 \Psi
\right]
\nonumber \\
& =
-\frac{a'}{8}
\left[
\bar{\Psi}({\g}^0 s^{01} + s^{01} {\g}^0) \Psi 
-\bar{\Psi}({\g}^1 \g^0\g^k s^{0k} +s^{0k}\gamma^k\gamma^0{\g}^1) \Psi
\right]
\nonumber \\
& =
-\frac{a'}{8}
{\psi}\adj \bar{q} 
\left(
  {\g}^0 s^{01} + s^{01} {\g}^0
- {\g}^1 \g^0\g^k s^{0k} - s^{0k}\gamma^k\gamma^0{\g}^1
\right)q \psi
\nonumber \\
& =
-\frac{a'}{8}
\int dq_1dq_2
{\psi}\adj 
\left(-4(q_1^2 - q_2^2 - q_3^2 + q_4^2)(q_1^2 + q_2^2 + q_3^2 + q_4^2)\right) 
\psi,
\\
\label{eq:T02calculation}
T_{02} 
& =
-\frac{a'}{8}
{\psi}\adj \bar{q} 
\left(
  {\g}^0 s^{02} + s^{02} {\g}^0
- {\g}^2 \g^0\g^k s^{0k} - s^{0k}\gamma^k\gamma^0{\g}^2
\right)q \psi
\nonumber \\
&=
 -\frac{a'}{8}
\int dq_1dq_2
{\psi}\adj 
\left(8(q_1q_2 - q_3q_4)(q_1^2 + q_2^2 + q_3^2 + q_4^2)\right) 
\psi,
\\
\label{eq:T03calculation}
T_{03} 
& =
-\frac{a'}{8}
{\psi}\adj \bar{q} 
\left(
  {\g}^0 s^{03} + s^{03} {\g}^0
- {\g}^3 \g^0\g^k s^{0k} - s^{0k}\gamma^k\gamma^0{\g}^3
\right)q \psi
\nonumber \\
&=
 -\frac{a'}{8}
\int dq_1dq_2
{\psi}\adj 
\left(-8(q_1q_3 + q_2q_4)(q_1^2 + q_2^2 + q_3^2 + q_4^2)\right) 
\psi,
\end{align}
and
\begin{align}
\label{eq:T12calculation}
T_{12} 
& =
\frac{ia}{4}
\bar{\Psi}
(
{\g}^1 \Omega_2 + \Omega_2 {\g}^1
+
{\g}^2 \Omega_1 + \Omega_1 {\g}^2
) 
\Psi 
\nonumber \\
& =
\frac{a'}{8}
{\psi}\adj \bar{q} 
\left(
  {\g}^1 s^{02} + s^{02} {\g}^1
+
  {\g}^2 s^{01} + s^{01} {\g}^2
\right)q \psi
\nonumber \\
&=
 \frac{a'}{8}
\int dq_1dq_2
{\psi}\adj 
\left(
- 4q_1^3q_4 - 12q_1^2q_2q_3 + 12q_1q_2^2q_4 
+ 12q_1q_3^2q_4 - 4q_1q_4^3 + 4q_2^3q_3 + 4q_2q_3^3 
- 12q_2q_3q_4^2
\right) 
\psi,
\\
T_{23}
&=
 \frac{a'}{8}
\int dq_1dq_2
{\psi}\adj 
\left(
4q_1^3q_2 - 12q_1^2q_3q_4 + 4q_1q_2^3 - 12q_1q_2q_3^2 
- 12q_1q_2q_4^2 - 12q_2^2q_3q_4 + 4q_3^3q_4 + 4q_3q_4^3
\right) 
\psi,
\\
T_{31}
&=
 \frac{a'}{8}
\int dq_1dq_2
{\psi}\adj 
\left(
- 2q_1^4 + 12q_1^2q_3^2 + 2q_2^4 
- 12q_2^2q_4^2 - 2q_3^4 + 2q_4^4
\right) 
\psi.
\end{align}
This directional asymmetry, coupled with the linear dependence 
on the Hubble constant, may provide an internally 
driven mechanism for second-order generation of anisotropy of the 
{\it initially isotropic} cosmic microwave background 
(via: ``new Dirac's field'' $\rightarrow$ 
``gravity'' $\rightarrow$ ``electromagnetic field''). 
To put it differently, an expanding universe populated by 
the new Dirac field has no choice but to evolve anisotropically.

\appendix

\section{Derivation of the stress-energy tensor}
\label{appendix1}

Our derivation parallels that for the usual Dirac field as presented 
in \cite{shapiro2022,buchbinder2021}. One may compare 
the appearance of the divergence term in our final 
result below with that given in Eq.\ (6.137) in Ref.\ \cite{dewitt2003}. 

Denoting 
$\d g_{\mu\nu} = h_{\mu \nu}$, we have,
\begin{align}
i \, \d {\cal L}_{\rm D}
&=
\d \left(
\sqrt{-g}
\left\{
\frac{1}{2}
\left[
\bar{\Psi} \breve{\gamma}^\mu (\nabla_\mu \Psi)
-
(\nabla_\mu \bar{\Psi}) \breve{\gamma}^\mu \Psi
\right]
-m \bar{\Psi} \Psi
\right\}
\right)
\nonumber \\
&=
(\d \sqrt{-g})
\left\{
\frac{1}{2}
\left[
\bar{\Psi} \breve{\gamma}^\mu (\nabla_\mu \Psi)
-
(\nabla_\mu \bar{\Psi}) \breve{\gamma}^\mu \Psi
\right]
-m \bar{\Psi} \Psi
\right\}
+
\sqrt{-g} \,
\d 
\left\{
\frac{1}{2}
\left[
\bar{\Psi} \breve{\gamma}^\mu (\nabla_\mu \Psi)
-
(\nabla_\mu \bar{\Psi}) \breve{\gamma}^\mu \Psi
\right]
\right\}
\nonumber \\
&=
\frac{1}{2} \,
h_{\mu\nu} \, \sqrt{-g} \, g^{\mu\nu}
\left\{
\frac{1}{2}
\left[
\bar{\Psi} \breve{\gamma}^\a (\nabla_\a \Psi)
-
(\nabla_\a \bar{\Psi}) \breve{\gamma}^\a \Psi
\right]
-m \bar{\Psi} \Psi
\right\}
+
\frac{1}{2} \, \sqrt{-g} \,
\d F,
\end{align}
where
\begin{align}
\d F
&=
\d 
\left[
\bar{\Psi} \breve{\gamma}^\mu (\nabla_\mu \Psi)
-
(\nabla_\mu \bar{\Psi}) \breve{\gamma}^\mu \Psi
\right]
\nonumber \\
&=
\d 
\left\{
e^\mu_a \, 
\left[
\bar{\Psi}  {\gamma}^a (\nabla_\mu \Psi)
-
(\nabla_\mu \bar{\Psi}) {\gamma}^a \Psi
\right]
\right\}
\nonumber \\
&=
(\d e^\mu_a)
\left[
\bar{\Psi}  {\gamma}^a (\nabla_\mu \Psi)
-
(\nabla_\mu \bar{\Psi}) {\gamma}^a \Psi
\right]
+
e^\mu_a \,
\d 
\left[
\bar{\Psi}  {\gamma}^a (\nabla_\mu \Psi)
-
(\nabla_\mu \bar{\Psi}) {\gamma}^a \Psi
\right]
\nonumber \\
&=
-\frac{1}{2}\, h^{\mu}_{\nu} \, e^\nu_a
\left[
\bar{\Psi}  {\gamma}^a (\nabla_\mu \Psi)
-
(\nabla_\mu \bar{\Psi}) {\gamma}^a \Psi
\right]
+
e^\mu_a \,
\d 
\left[
\bar{\Psi}  {\gamma}^a (\partial_\mu + \Omega_\mu) \Psi
-
\bar{\Psi}(\cev{\partial}_{\mu} - \Omega_{\mu}) {\gamma}^a \Psi
\right]
\nonumber \\
&=
-\frac{1}{2}\, h_{\mu \nu} 
\left[
\bar{\Psi} \breve{\gamma}^\nu (\nabla^\mu \Psi)
-
(\nabla^{\mu} \bar{\Psi}) \breve{\gamma}^\nu \Psi
\right]
+
e^\mu_a \,
\bar{\Psi} 
\left[ 
{\gamma}^a (\d \Omega_\mu) 
+
(\d \Omega_{\mu}) {\gamma}^a 
\right]
\Psi .
\end{align}
Now,
\begin{align}
e^\mu_a \,
\bar{\Psi} 
\left[ 
{\gamma}^a (\d \Omega_\mu) 
+
(\d \Omega_{\mu}) {\gamma}^a 
\right]
\Psi 
&=
-\frac{1}{4}\, (\d \omega_{ab\mu}) \, 
e^\mu_c \,
\bar{\Psi} 
\left[ 
{\gamma}^c (\sigma^{ab}+is^{ab}) 
+
(\sigma^{ab}+is^{ab}) {\gamma}^c 
\right]
\Psi 
\nonumber \\
&=
-\frac{1}{4}\, (\d \omega_{ab\mu}) \, 
e^\mu_c \,
\bar{\Psi} 
\left(
{\gamma}^c \sigma^{ab}
+
\sigma^{ab} {\gamma}^c 
\right)
\Psi 
-\frac{i}{4}\, (\d \omega_{ab\mu}) \, 
e^\mu_c \,
\bar{\Psi} 
\left(
{\gamma}^c s^{ab}
+
s^{ab}{\gamma}^c 
\right)
\Psi.
\end{align}
Using
\begin{align}
\quad
s^{ab}q=qs^{ab}+i \sigma^{ab}q,
\quad
\bar{q}s^{ab}=s^{ab}\bar{q}+i\bar{q}\sigma^{ab},
\quad
s^{ab}=\frac{i}{2}[g^{a},g^{b}],
\quad
g^{a}\equiv \frac{1}{2}(\bar{q} \gamma^{a} q),
\end{align}
we get,
\begin{align}
-i\bar{\Psi} 
\left( 
{\gamma}^c s^{ab}
+
s^{ab}{\gamma}^c 
\right)
\Psi
&=
-i{\psi}\adj \bar{q} 
\left( 
{\gamma}^c s^{ab}
+
s^{ab}{\gamma}^c 
\right)
q\psi
\nonumber \\
&=
-i{\psi}\adj 
\left[
\bar{q} {\gamma}^c \left( q s^{ab} +i \sigma^{ab}q \right)
+
\left(  s^{ab}\bar{q} +i \bar{q}\sigma^{ab} \right){\gamma}^c q
\right]
\psi
\nonumber \\
&=
{\psi}\adj 
\left(
\bar{q} {\gamma}^c  \sigma^{ab}q 
+
\bar{q}\sigma^{ab} {\gamma}^c q
\right)
\psi
-
i{\psi}\adj 
\left(
\bar{q} {\gamma}^c  q s^{ab} 
+
 s^{ab}\bar{q}{\gamma}^c q
\right)
\psi
\nonumber \\
&=
\bar{\Psi} 
\left(
{\gamma}^c  \sigma^{ab}
+
\sigma^{ab} {\gamma}^c 
\right)
\Psi
+
{\psi}\adj 
\left(
g^c [g^a,g^b]
+
[g^a,g^b] g^c
\right)
\psi,
\end{align}
and thus,
\begin{align}
e^\mu_a \,
\bar{\Psi} 
\left[ 
{\gamma}^a (\d \Omega_\mu) 
+
(\d \Omega_{\mu}) {\gamma}^a 
\right]
\Psi 
&=
\frac{1}{4}\, (\d \omega_{ab\mu}) \, e^{\mu}_{c}\, 
{\psi}\adj 
\left(
g^c [g^a,g^b]
+
[g^a,g^b] g^c
\right)
\psi.
\end{align}
Next, taking into account that
\BEq
(\d \omega_{ab\mu})
=
(e_a^{\tau}e_b^{\lambda}-e_b^{\tau}e_a^{\lambda})
\nabla_{\lambda}h_{\mu\tau},
\EEq
we find, dropping a total divergence,
\begin{align}
e^\mu_a \,
\bar{\Psi} 
\left[ 
{\gamma}^a (\d \Omega_\mu) 
+
(\d \Omega_{\mu}) {\gamma}^a 
\right]
\Psi 
&=
-\frac{1}{4}\,
h_{\mu\nu} \,
(e_a^{\nu}e_b^{\lambda}-e_b^{\nu}e_a^{\lambda})
\, e^{\mu}_{c}\, 
\nabla_{\lambda}
\left\{
{\psi}\adj 
\left(
g^c [g^a,g^b]
+
[g^a,g^b] g^c
\right)
\psi
\right\}
\nonumber \\
&=
-\frac{1}{2}\,
h_{\mu\nu} \,
e^{\mu}_{c} e_a^{\nu}e_b^{\lambda}\, 
\nabla_{\lambda}
\left\{
{\psi}\adj 
\left(
g^c [g^a,g^b]
+
[g^a,g^b] g^c
\right)
\psi
\right\}
\nonumber \\
&=
-\frac{1}{2}\,
h_{\mu\nu} \,
\nabla_{\lambda}
\left\{
{\psi}\adj 
\left(
g^\mu [g^\nu,g^\lambda]
+
[g^\nu,g^\lambda] g^\mu
\right)
\psi
\right\}
\nonumber \\
&=
-\frac{1}{4}\,
h_{\mu\nu} \,
\nabla_{\lambda}
\left\{
{\psi}\adj 
\left[
\left(g^\mu g^\nu+g^\nu g^\mu\right)g^\lambda
-
g^\lambda \left(g^\mu g^\nu+g^\nu g^\mu\right)
\right]
\psi
\right\},
\end{align}
which gives, upon symmetrizing,
\begin{align}
\d F
&=
-\frac{1}{2}\, h_{\mu \nu} 
\left(
\bar{\Psi} \breve{\gamma}^{(\mu} \nabla^{\nu)} \Psi
-
\nabla^{(\mu} \bar{\Psi}\breve{\gamma}^{\nu)} \Psi
+
\frac{1}{2}\,
\nabla_{\lambda}
\left\{
{\psi}\adj 
\left[
\left(g^\mu g^\nu+g^\nu g^\mu\right)g^\lambda
-
g^\lambda \left(g^\mu g^\nu+g^\nu g^\mu\right)
\right]
\psi
\right\}
\right).
\end{align}
Combining all these results produces the stress-energy tensor,
\begin{align}
\label{eq:SETderived}
T^{\mu\nu}
&= 
-\frac{2}{\sqrt{-g}}\frac{\delta S_{\rm D}}{\delta g_{\mu\nu}}
\nonumber \\
&=
-i \, 
\left(
\frac{1}{2}
\left[
\bar{\Psi} \breve{\gamma}^{(\mu} \nabla^{\nu)} \Psi
-
\nabla^{(\mu} \bar{\Psi} \breve{\gamma}^{\nu )} \Psi
\right]
-
\frac{1}{2}g^{\mu\nu}
\left[
\bar{\Psi} \breve{\gamma}^{\alpha} \nabla_{\alpha} \Psi
-
\nabla_{\alpha} \bar{\Psi} \breve{\gamma}^{\alpha} \Psi
\right]
+m g^{\mu\nu}\bar{\Psi} \Psi
\right)
\nonumber \\
&\quad
-
\frac{i}{2}\,
\nabla_{\lambda}
\left[
{\psi}\adj 
\left\{
\left(g^\mu g^\nu+g^\nu g^\mu\right)g^\lambda
-
g^\lambda \left(g^\mu g^\nu+g^\nu g^\mu\right)
\right\}
\psi
\right].
\end{align}
Notice that in the classical limit, when noncommutativity of $q$'s can be  ignored, the combination of $g$'s in the curly brackets in Eq.\ (\ref{eq:SETderived}) vanishes, which allows us to drop the divergence term.

\end{document}